# Identifying recombination hotspots using population genetic data

Adam Auton[1,*], Simon Myers[2] and Gil McVean[2]

[1]Department of Genetics, Albert Einstein College of Medicine, 1301 Morris Park Ave., Bronx, NY 10461, USA.
[2]Wellcome Trust Centre for Human Genetics, University of Oxford, Roosevelt Drive, Oxford OX3 7BN, UK.

## ABSTRACT

**Motivation:** Recombination rates vary considerably at the fine scale within mammalian genomes, with the majority of recombination occurring within hotspots of ~2 kb in width. We present a method for inferring the location of recombination hotspots from patterns of linkage disequilibrium within samples of population genetic data.

**Results:** Using simulations, we show that our method has hotspot detection power of approximately 50-60%, but depending on the magnitude of the hotspot. The false positive rate is between 0.24 and 0.56 false positives per Mb for data typical of humans.

**Availability:** http://github.com/auton1/LDhot

**Contact:** adam.auton@einstein.yu.edu

## 1  INTRODUCTION

The majority of recombination in mammalian genomes occurs in highly localized regions known as recombination hotspots. These hotspots are approximately 2 kb in width, have recombination rates that are tens or thousands of times greater than their surrounding regions, and are a ubiquitous feature of mammalian genomes (Myers, et al., 2005). We have previously published a computational method for estimating recombination rates using samples of population genetic data. This method, known as *LDhat*, exploits patterns of linkage disequilibrium and can accurately estimate recombination rates at both fine and broad scales (McVean, et al., 2004).

Potential recombination hotspots appear in the output of *LDhat* as localized peaks within the recombination rate estimates. However, while such peaks often represent true variation in the underlying recombination rate, noise in the estimator means some are expected to be false positives. To assess the significance of the observed peaks, we have previously applied a method known as *LDhot* (Myers, et al., 2005). This method uses extensive coalescent simulations to assess the significance of the observed recombination peak, and hence formally test for the presence of a hotspot. In humans, *LDhot* has been used to identify approximately 30,000 hotspots across the genome, and was central in the discovery of a DNA motif associated with recombination hotspots (Myers, et al., 2008). This DNA motif was subsequently identified as a binding site of the zinc-finger protein, PRDM9, which is now believed to be responsible for localizing the vast majority of hotspots in humans and mice (Baudat, et al., 2010; Berg, et al., 2010).

However, the initial implementations of *LDhot* did not consist of coherent package, but rather consisted of a collection of *ad hoc* scripts. In addition, the package required use of a large computing cluster even for small datasets, and had an unwieldy number of parameters that needed to be tuned to a specific dataset. To overcome these limitations, we now describe a self-contained version of *LDhot* that avoids many of these issues and is available for public download.

## 2  APPROACH

The level of linkage disequilibrium (LD) between loci is informative of the amount of historical recombination between them, as higher recombination rates tend to break down the correlation more rapidly over successive generations. In a formal sense, the LD patterns are determined by the (unknown) historical genealogy of the data. The distribution of these genealogies is often modeled using the coalescent with recombination, which has the (varying) recombination rate as a parameter. While inference about the recombination parameters would be straightforward given the genealogy, the genealogy is generally unknown and so must be treated as missing data. To obtain the true likelihood of the data given a set of parameters, one would like to be able to integrate over all possible genealogical histories of the sample. However, even using modern computational approaches, this is usually impractical as the number of possible genealogies that contribute to the likelihood is infeasibly large for all but the smallest of datasets.

To overcome this, *LDhat* uses an approximation to the likelihood obtained by treating all pairs of sites independently. Let the phased polymorphism data for a pair of sites $i$ and $j$ be $X_{ij}$. The two sites are separated by a population map distance, $\rho_{ij} = 4N_e r_{ij}$. The likelihood of the recombination rate given just the pair of sites, $L(\rho_{ij}|X_{ij})$, can be estimated under a coalescent model using numerical methods such as importance sampling (Fearnhead and Donnelly, 2001). The calculation is still computationally expensive, but can be performed once in advance, with the results being stored in a lookup table for later use.

If $\boldsymbol{\rho}$ represents the recombination rate profile, and $\boldsymbol{X}$ represents the complete dataset, then the 'composite likelihood' of the complete dataset can be calculated as the product over all pairs of sites within some arbitrary distance of each other (in *LDhat*, we default to 50 SNPs):

$$CL(\boldsymbol{\rho}|\boldsymbol{X}) = \prod_{1 \le |i-j| \le 50} L(\rho_{ij}|X_{ij})$$

For a given number of sequences, the pairwise likelihoods can be pre-computed for all possible haplotype configurations consisting of two sites. This pre-computation allows the composite likelihood itself to be calculated quickly for a given dataset and arbitrary $\boldsymbol{\rho}$, and this enables exploration of the space of possible recombination rates. In *LDhat*, a reversible jump MCMC scheme is used to explore the space of possible piecewise-constant recombination rate profiles (McVean, et al., 2004).

A limitation of the composite likelihood is that its surface tends to be overly peaked relative to the true likelihood, and it therefore cannot be easily used to assess the uncertainty around a point estimate. To formally test for the presence of a hotspot, we compare a model in which the recombination rate is constant across a window (the null model), to a model in which the recombination rate within a small window at the center of the region is allowed to differ from the surroundings (the alternative model). In brief, our approach performs a likelihood ratio test, using coalescent simulations to determine the null distribution in the absence of a hotspot. Simulations are required because the approximations used in the likelihood mean the likelihood ratio test statistic is not expected to have a standard distribution (e.g. a chi-squared distribution) under the null.

---

*To whom correspondence should be addressed.

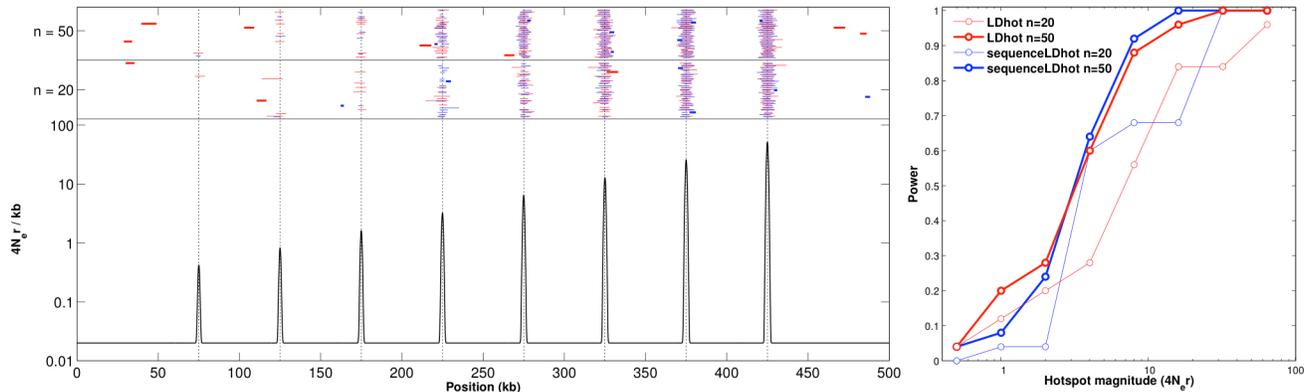

**Figure 1. The left panel shows the simulated recombination rate on a log scale, covering 500kb with eight hotspots of varying magnitude. The positions of hotspot calls are shown above for the simulations with 20 and 50 chromosomes, with *LDhot* calls shown in red and *sequenceLDhot* shown in blue. False positive calls are in bold. The right panel shows the estimated power as a function of hotspot magnitude on a log scale.**

Specifically, the test statistic is the ratio of the composite likelihood maximized under the null model, to that maximized under the alternative model. Let the null model be specified by a single constant recombination rate, $\rho_{const}$. Likewise, let the alternative model be specified by two recombination rates, $\rho_{hot}$ and $\rho_{bg}$, representing the recombination rate within the hotspot and the background rate respectively. The test statistic is:

$$\Lambda = -2\ln\left(\frac{\sup[CL(\rho_{hot}, \rho_{bg}|\mathbf{X})]}{\sup[CL(\rho_{const}|\mathbf{X})]}\right)$$

Earlier versions of *LDhot* used extensive coalescent simulations to estimate the test statistic for a variety of recombination rates and SNP densities, and stored the results in lookup-tables to allow for efficient hotspot detection on a genome-wide basis. In contrast, our new method performs simulations 'on the fly' for each putative hotspot which, while computationally more expensive to apply on a genome-wide basis, is more convenient to distribute in a unified package and allows for more flexibly in simulation parameters.

To test for the presence of a recombination hotspot, our method starts by constructing overlapping 3 kb windows across the region that are separated by 1kb. Rather than testing all possible windows for the presence of a hotspot, we only consider windows that overlap local recombination rate maxima within the *LDhat* rate estimates. For each tested window, we define the background region as ± 50 kb of the window center. The window sizes described here are the default parameters, which can be altered by the user if necessary.

In general, we perform up to 1,000 simulations for each putative hotspot, although simulations are cut short if it is clear that there is no evidence of significance. Simulations are conducted under the neutral coalescent with recombination and assuming an infinite sites mutation model. The simulations use a constant recombination rate across the region, drawn from an exponential distribution with mean equal to the average rate estimated by *LDhat* for the real data.

The locations of the simulated mutations are set to match the real data. In addition, in order to match the simulated data more closely, we also attempt to match the simulated minor allele count at each locus with that of the real data. There is no obvious method by which this may be achieved exactly in the coalescent simulation framework, and hence we use an *ad hoc* heuristic approach. Let the minor allele count in the real data be $i$, and $j$ be the minor allele count that would result from placing a mutation on branch $b$ in the simulated genealogy. We assume $i$ and $j$ represent successive samples from the same population, and estimate $P(j|i)$ using a beta-binomial model with uniform prior. When placing mutations on the simulated coalescent genealogy, we weight the probability of a mutation occurring on a given branch by $L_b P(j|i)$, where $L_b$ is the length of the branch. While this approach is heuristic, it provides a convenient means to improve the match between the real and simulated data, and thus potentially improve power to detect a real hotspot for a given type I error. The frequency matching can be disabled by the user if required.

Having performed $N$ simulations for a given window, an empirical p-value is estimated as $(s + 1)/(N + 1)$, where $s$ is the number of simulations with a test statistic more extreme than that obtained from the real data. For windows in which the data test statistic is in the extreme tail of the null distribution (i.e. $s < 10$ with $N \geq 1000$), we apply the method of Knijnenburg *et al.* (2009) to fit a Generalized Pareto Distribution to the null distribution tails, which allows estimation of an approximated p-value that takes into account the deviation of the test statistic from the null distribution.

We call a hotspot if a window has a p-value less than a specified threshold, generally taken as 0.001. To identify the boundaries of a hotspot, we combine any adjacent windows that achieve significance at a lower threshold such as 0.01, and expand the resulting region out to the nearest polymorphism on either side.

## 3 RESULTS

In order to assess the power and false positive rate of our method, we conducted two separate simulation studies. We used coalescent simulations to simulate 500 kb regions for populations of constant past population size. The population mutation rate, $\theta = 4N\mu$, was set to 1 per kb. For each simulated dataset, we used the *interval* program from *LDhat* with a block penalty of 5 to estimate recombination rates. These estimates were passed into the hotspot detection method. We compared our method to that of *sequenceLDhot* (Fearnhead, 2006), which was given the true background rate as an input parameter.

### 3.1 Uniform recombination rate

We simulated 25 datasets with 20 chromosomes a constant recombination crossover rate of $\rho = 0.44$ per kb, equivalent to 1.1 cM/Mb assuming $N_e = 10,000$, and consistent with broad-scale crossover estimates in humans (Kong, et al., 2010). In total, 5 hotspots were called with p < 0.001, which implies a false positive rate of 1 hotspot / 2.5 Mb. The average run time for each dataset was 37 minutes on cluster nodes with CPU speeds ranging from 1.6 to 2.6 Ghz (and not counting the *LDhat* runtime). In contrast, *sequenceLDhot* called 1 false positive and ran in an average of 6 minutes on similar CPUs.

### 3.2 Recombination hotspots

We simulated 25 datasets with 20 chromosomes, each containing eight hotspots spaced 50kb apart (Figure 1). The eight hotspots were 2 kb in width and had magnitudes of $\rho = 0.5, 1, 2, 4, 8, 16, 32,$ and $64$, equivalent to $1.25 \times 10^{-3}$ cM for the smallest hotspot, and $0.32$ cM for the largest. The background recombination rate was $\rho = 0.02$ per kb, equivalent to $0.05$ cM/Mb. These parameters were chosen to approximately match data obtained from sperm typing experiments in humans (e.g. Jeffreys, et al., 2005). We note that median magnitude of detected hotspots in the human genome is $0.043$ cM (International HapMap Consortium, 2007), corresponding to $\rho = 17.2$.

We called a true positive if the called region overlapped with one of the simulated hotspots. At the 0.001 significance level, the power to call the eight hotspots was 4%, 12%, 20%, 28%, 56%, 84%, 84%, and 96% respectively. In total, there were 3 false positives, equivalent to approximately 1 false positive per 4 Mb of sequence. The average run time was 64 minutes. In contrast, *sequenceLDhot* ran in an average of 6 minutes, called 6 false positives, and had power to detect the eight hotspots of 0%, 4%, 4%, 60%, 68%, 68%, 100% and 100% respectively.

We repeated the simulations with a larger sample size of 50 chromosomes. In this case, the power to call the eight hotspots was 4%, 20%, 28%, 60%, 88%, 96%, 100%, and 100% respectively. There were 7 false positives, corresponding to approximately 1 per 1.8 Mb. The average run time was 36 minutes. Using the same data, *sequenceLDhot* ran in an average of 14 minutes, produced 7 false positives, and had 4%, 8%, 24%, 64%, 92%, 100%, 100%, and 100% detection power for the eight hotspots.

Across all the hotspot simulations, *LDhot* identified 55% of the simulated hotspots with 10 false positives, whereas *sequenceLDhot* identified 56% with 13 false positives. None of the *LDhot* false positives had estimated magnitudes of $\rho < 5$ (~0.0125 cM), suggesting that exclusion of hotspots with small magnitudes could be used to reduce the false positive rate. Conversely, most *sequenceLDhot* false positives are in the vicinity of true hotspots, suggesting that they may be largely due to mislocalization.

## 4    CONCLUSIONS

Our simulations would suggest we have ~50-60% power to detect hotspots of a moderate magnitude, similar to previously reported estimates for *LDhot* (Myers, et al., 2005). However, we also found that it is challenging to detect hotspots that are relatively weak at the population level. This may be a particular issue in populations and species with "flatter" recombination profiles – for example, populations of mammals with higher PRDM9 diversity, or species with different recombination biology to that of humans.

A major advantage of this new implementation is that it is a self-contained program that can be used in conjunction with the *LDhat* package. As such, it is simple to run and provides a convenient means to assess the evidence for recombination hotspots within population genetic datasets. However, we note that the method is still expensive in terms of computational time, and hence a large compute cluster is required to run on a genome-wide basis.